\documentclass[twocolumn,pra,showpacs]{revtex4}

\usepackage{graphicx}
\usepackage{rotating}
\usepackage{amsmath}
\usepackage{amsfonts}
\usepackage{amssymb}
\usepackage{enumerate}
\usepackage{longtable}
\usepackage{dcolumn}
\usepackage{bm}

\begin{document}

\title{Temperature in One-Dimensional Bosonic Mott Insulators}

\author{A. Reischl$^1$}
\author{K.P. Schmidt$^2$}
\email{kaiphillip.schmidt@epfl.ch}
\author{G.S. Uhrig$^3$}
\affiliation{$^1$ Institut f\"ur Theoretische Physik, Universit\"at zu K\"oln, Z\"ulpicher Stra{\ss}e 77, 50937 K\"oln, Germany}
\affiliation{$^2$ Institute of Theoretical Physics, \'{E}cole Polytechnique F\'{e}d\'{e}rale de Lausanne, CH 1015 Lausanne, Switzerland}
\affiliation{$^3$ Theoretische Physik, Geb. 38, FR 7.1, 
Universit\"at des Saarlandes, D-66123 Saarbr\"ucken, Germany}
\date{\rm\today}

\begin{abstract} 
The Mott insulating phase of a one-dimensional bosonic  gas
trapped in optical lattices is described by a Bose-Hubbard model. 
A continuous unitary transformation is used to map
this model onto an effective model conserving the number of elementary
excitations. We obtain quantitative results for
the kinetics and for the spectral weights of the low-energy excitations 
for a broad range of parameters in the insulating phase. 
By these results, recent Bragg spectroscopy experiments
are explained. Evidence for a significant temperature of the order
of the microscopic energy scales is found.
\end{abstract}



\maketitle

\section{Introduction}
Ultracold atoms in optical lattices provide tunable experimental
realizations of correlated many-particle systems. Even the dimensionality
of these systems can be chosen between one and three.
Recently, the transition between superfluid and Mott insulating (MI) phases
has been observed by tuning the ratio between the kinetic and the interaction
energy \cite{grein02,pared04}. Dynamical aspects can also be addressed 
 experimentally \cite{stofe04,kohl05}.

There are many theoretical treatments of these issues, ranging from 
mean-field approaches \cite{jaksc98} to more sophisticated 
techniques \cite{fishe89,pai96,kashu96,kuhne98,elstn99,kuhne00}. 
The transition at zero temperature ($T=0$)
has been analyzed in one-dimensional 
systems \cite{pai96,kashu96,kuhne98,elstn99,kuhne00}. 
But there is still a multitude of open questions.
Our aim is to explain the dynamics of the excitations
 which has moved in the focus recently
\cite{stofe04,kohl05,batro05}.

The present work is intended to clarify the significance of various
excitation processes depending on the  kinetic and the 
interaction energy and on the temperature. To this end, we will
analyze the spectral weights of the
one-dimensional MI phase with one boson per site $n=1$.

The low-energy excitations are either double occupancies ($n=2$),
which we will call `particles', or they are holes ($n=0$). Both excitations 
are gapped as long as the system is insulating. They become soft at the 
quantum critical point which separates the insulator from the superfluid.
We use a continuous unitary transformation (CUT)
\cite{wegne94,knett00a} to obtain an effective Hamiltonian 
in terms of the elementary excitations `particle' and `hole'. 
This effective Hamiltonian
conserves the number of particles and of holes \cite{knett03a}.
The CUT is realized in real space in close analogy to the 
derivation of the generalized $t$-$J$ model from the Hubbard model
\cite{reisc04}. The strongly correlated many-boson problem
is reduced to a problem involving a small number of particles and holes.
This simplification enables us to calculate 
kinetic properties like the dispersions  and
spectral properties like spectral weights.

The article is set-up as follows. First, the model and the relevant
observable are introduced (Sect.\ II). 
Then, the method used is presented and described (Sect.\ III).
The spectral weights are computed in Sect.\ IV. In Sect.\ V, the
experiment is analyzed which requires calculations at finite temperatures
as well. Finally, the results are discussed in Sect.\ VI.

\section{Model}
To be specific, we study the Bose-Hubbard model 
$H = t H_t+ U H_U$ in one dimension
\begin{equation}
\label{hamiltonian}
 H = -t\sum\limits_i( b^\dagger_i b^{\phantom{\dagger}}_{i+1}+b^\dagger_{i+1}
 b^{\phantom{\dagger}}_i ) + (U/2)\sum_i \hat{n}_i ( \hat{n}_i-1) 
\end{equation}
where the first term is the kinetic part $t H_t$ and the second term the 
repulsive interaction $U H_U$ with $U>0$. The 
 bosonic annihilation (creation) operators are denoted by 
$b^{(\dagger)}_i$, the number of bosons by 
$\hat{n}_i =b^\dagger_i b^{\phantom{\dagger}}_i$.
If needed the term $H_\mu= -\mu\sum_i \hat{n}_i$ is added to $H$
to control the particle number.
For numerical simplicity, we  truncate the local bosonic Hilbert space 
to four states. This does not change the relevant physics significantly
\cite{pai96,kashu96,kuhne98}.

Besides the Hamiltonian we need to specify the excitation operator $R$. In the
set-up of Refs.~\onlinecite{stofe04} and \onlinecite{kohl05} the depth
of the optical lattices is changed periodically to excite the system. 
In terms of the tight binding model (\ref{hamiltonian}) this amounts
to a periodic change of $t$ and of $U$
leading to 
\begin{equation}
R \propto \delta t H_t +\delta U H_U\ .
\end{equation} 
Since multiples of $H$ do not induce excitations we consider
\begin{equation}
R \to \tilde R = R- (\delta U/U) H\ .
\end{equation} 
Both operators $R$ and $\tilde R$ induce the same transitions
$\langle n | R | m \rangle = \langle n | \tilde R | m \rangle$
where $| n \rangle \neq | m \rangle$ are eigen states of $H$.
Eventually, the relevant part of $R$ is proportional to $H_t$.
For simplicity, we set the factor of proportionality to one.

If the interaction dominates ($U/t \to \infty$) the ground state of
(\ref{hamiltonian}) is the product state of precisely one boson per site
$|\text{ref}\rangle=|1\rangle_1\otimes |1\rangle_2 \ldots\otimes |1\rangle_N$,
where $|n\rangle_i$ denotes the local state at site $i$ with $n$ bosons.
We take $|\text{ref}\rangle$  as our reference state; all deviations
from $|\text{ref}\rangle$ are considered as elementary excitations. 
Since we restrict
our calculation to four states per site we define three creation operators:
$h^\dagger_i |1\rangle_i= |0\rangle_i$ induces a hole at site $i$,
$p^\dagger_i |1\rangle_i= |2\rangle_i$ induces a particle at site $i$,
and $d^\dagger_i |1\rangle_i= |3\rangle_i$ induces a double-particle
at site $i$. The operators $h$, $p$ and $d$ obey the 
commutation relations for hardcore bosons.

\section{CUT}
The Hamiltonian (\ref{hamiltonian}) conserves the number of 
bosons $b$. But if it is rewritten in terms of $h,p$, and $d$ it is no longer
particle-conserving, e.g., the application of $H_t$ to $|\text{ref}\rangle$
generates a particle-hole pair. We use a CUT defined by
\begin{equation}
  \label{eq:fleq}
  \partial_l H (l) = [\eta (l),H(l)]
\end{equation}
to transform $H(l=0)=H$ from its form in 
(\ref{hamiltonian}) to an effective Hamiltonian
$H_\text{eff}:=H(l=\infty)$ which \emph{conserves} the number
of elementary excitations, i.e., $[H_U,H_\text{eff}]=0$. An appropriate
choice of the infinitesimal generator $\eta$ is defined by the matrix
elements  
\begin{equation}
\label{eq:generator}
\eta_{i,j} (l)={\rm sgn}\left(q_i-q_j\right)H_{i,j}(l)
\end{equation}
in an eigen basis of $H_U$; $q_i$ is the corresponding eigenvalue
\cite{knett00a}. The 
structure of the resulting $H_\text{eff}$ and the
implementation of the flow equation in second quantization in real space
is described in detail in Refs.\ \onlinecite{knett03a} and 
\onlinecite{reisc04}. 

\begin{figure}[t]
    \begin{center}
     \includegraphics[width=\columnwidth]{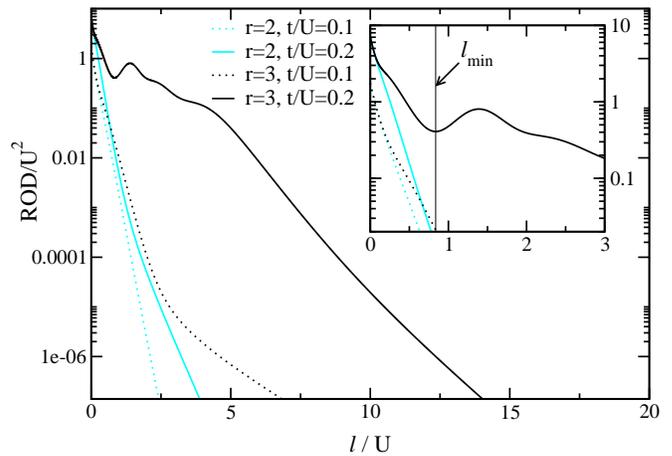}
    \end{center}
    \caption{Residual off-diagonality for values of $t/U\in \{0.1,0.2\}$ for
      maximal $r=2$ and $r=3$. {\em Inset:} Magnification for small values of
      the flow parameter $l$. The ROD for $r=3$ and $t/U=0.2$
      displays non-monotonic behavior. In this case, we stop 
      the flow at the first minimum of the ROD. 
      Its position is indicated by a vertical line.}
    \label{fig_ROD}
\end{figure}
The flow equation (\ref{eq:fleq}) generates a 
proliferating number of terms.
For increasing $l$, more and more hardcore
bosons are involved, e.g.,  annihilation and creation of three bosons 
$p^\dagger_i p^\dagger_{i+1} p^\dagger_{i+2}
p^{\phantom\dagger}_i p^{\phantom\dagger}_{i+1} p^{\phantom\dagger}_{i+2}$, 
and processes over a larger and larger range occur,
e.g., hopping over $j$ sites $p^\dagger_{i+j} p^{\phantom\dagger}_i$.
To keep the number of terms finite we omit normal-ordered terms
beyond a certain extension $r$.
Normal-ordering means that the creation operators of the elementary
excitations appear to the  left of the annihilation operators. If
terms appear which are not normal-ordered the commutation relations
are applied to rewrite them in normal-ordered form. The normal-ordering
is important since it ensures that only less important terms are omitted.
Generically, such terms involve more particles \cite{knett03a,reisc04}.

We define the extension $r$ as the distance between
the rightmost and the leftmost creation or annihilation
operator in a term \cite{knett03a,reisc04}. The extension $r$ of a term 
measures the range of the physical process which is described by
this term. So our restriction to a finite extension restricts the
range of processes kept in the description. This is the most
serious restriction in our approach.
Note that the extension $r$ implies for hardcore bosons
that at maximum $r+1$ bosons are involved.

For $l<\infty$, $H(l)$ still contains terms
like $p^\dagger_i h^\dagger_{i+1}$ which do not conserve the number
of particles. To measure the extent to which $H(l)$ deviates from the
desired particle-conserving form, we introduce the
residual off-diagonality (ROD) as the sum of the 
moduli squared of the coefficients of all terms which change the number of
elementary excitations.  The ROD for maximal extensions $r=2$ and $r=3$ is 
shown in Fig.~\ref{fig_ROD}. 
Ideally, the ROD decreases monotonically with $l$. 
The calculation with $r=2$ indeed shows the desired
monotonic behavior of the ROD for all values of $t/U$.

But non-monotonic behavior can also occur \cite{reisc04}. It is observed in 
Fig.\ \ref{fig_ROD} for extension $r=3$. 
While for small $t/U$ the ROD still decays monotonically it displays
non-monotonic behavior for larger $t/U$, see e.g.\ the ROD for $t/U=0.2$ and 
$r=3$ in Fig.~\ref{fig_ROD}. 
An uprise at larger values $l \gtrapprox 1/t$ signals that the 
intended transformation cannot be performed in a well-controlled way.
The uprise stems from matrix elements which do not decrease but increase,
at least in an intermediate interval of the running variable $l$.
Such an increase occurs if the number of excitations is not correlated
with the total energy of the states. This means that two states are linked by
an off-diagonal matrix element of which the state with low
number of excitations is \emph{higher} in total energy than the state with
a higher number of excitations. The increase of such a matrix element
bears the risk that the unavoidable
truncations imply a too severe approximation.

The situation that the number of excitations is not correlated with the
total energy can occur where the lower band edge of a continuum of more 
particles falls below the upper band edge of a continuum of less particles or
below the dispersion of a single excitation,
see, e.g., Ref.\ \onlinecite{schmi05b}. This situation implies additional
life-time effects since the states with less particles can decay
into the states with more particles. Even if the decay rates are very
small the CUT can be spoilt by them because the CUT in its form defined 
by Eq.\ (\ref{eq:generator}) correlates the particle number and the energy of
the states.

In order to proceed in a controlled way, we neglect the small life-time 
effects if they occur at all. This is done by stopping the CUT at
$l_\text{min}$ at the first minimum of the ROD. 
The position of $l_\text{min}$ for $t/U=0.2$ and $r=3$ is indicated in the
inset of Fig.~\ref{fig_ROD} by a vertical line. It is found that the remaining 
values of the ROD are small and thus negligible. Hence we omit the remaining
off-diagonal terms. 
Only close to the critical point, where
the MI phase vanishes, the present approach
becomes insufficient.

\begin{figure}[thbp]
    \begin{center}
     \includegraphics[width=\columnwidth,height=\columnwidth]
		     {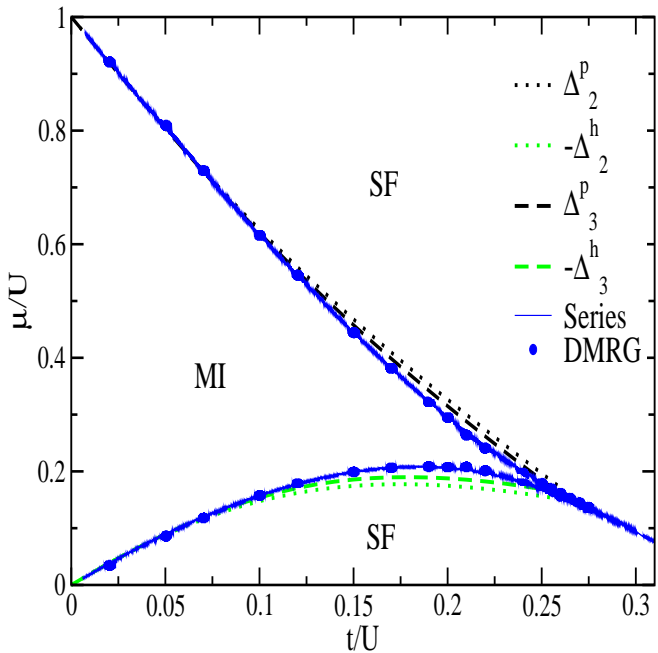}
    \end{center}
    \caption{(color online) Phase diagram of the Mott insulating (MI) and the
      superfluid (SF) phase in the $(t/U,\mu/U)$ plane.
      Dotted (dashed) lines show CUT results for maximal extension $r=2$ 
      ($r=3$). Solid lines (symbols) show series (DMRG) results from Refs.\
      \onlinecite{elstn99} and \onlinecite{kuhne98}.}
    \label{fig_PD}
\end{figure}
We check the reliability of our approach by comparing its results to those
of other methods
for the phase diagram in Fig.~\ref{fig_PD}. The upper
boundary of the MI phase is given by the particle gap
$\Delta^\text{p}:=\text{min}_k \omega^\text{p}(k)$. 
The lower boundary is  given by the negative hole gap
$-\Delta^\text{h}$ where
$\Delta^\text{h}:=\text{min}_k \omega^\text{h}(k)$.
The dotted (dashed) curves result from CUTs for $r=2$ ($r=3$).
For $r=2$, the CUT can be performed till $l=\infty$;
for $r=3$, the flow is stopped at $l_\text{min}$.
Solid curves depict the findings by series
expansion, the symbols those obtained by DMRG \cite{elstn99,kuhne98}. 
The agreement is very good in view of the truncation of the 
Hilbert space and in view of the low value of $r$. Note that the $r=3$
result agrees better with the series and DMRG results
than the $r=2$ result. As expected, the deviations
increase for larger values of $t$ because longer-range processes become
more important. Yet the values obtained by the CUT for the critical ratio 
$x_c:=t/U$, where the MI phase vanishes, are reasonable.
We find $x_c^{(r=2)}=0.271$ and $x_c^{(r=3)}=0.258$. By high accuracy 
density-matrix renormalization $x_c=0.297\pm0.01$ was found, see
Ref.\ \onlinecite{kuhne00} and references therein. Series expansion
provides $x_c=0.26\pm0.01$ which is very close to our  value $x_c^{(r=3)}$.
This fact underlines the similarity between series expansions and the
real space CUT as employed in the present work.

We conclude from the above findings for the phase diagram (Fig.\
\ref{fig_PD}) that the mapping to the particle-conserving $H_\text{eff}$
works very well for large parts of the phase diagram. It
does not, however, capture the Kosterlitz-Thouless nature of the transition
itself \cite{kuhne00}.
The CUT yields reliable results within the MI phase for 
$t\lessapprox 0.2U$. Henceforth, results  for this regime
will be shown which were obtained in the $r=3$ scheme.

\begin{figure}[t]
    \begin{center}
     \includegraphics[width=\columnwidth,height=\columnwidth]
		     {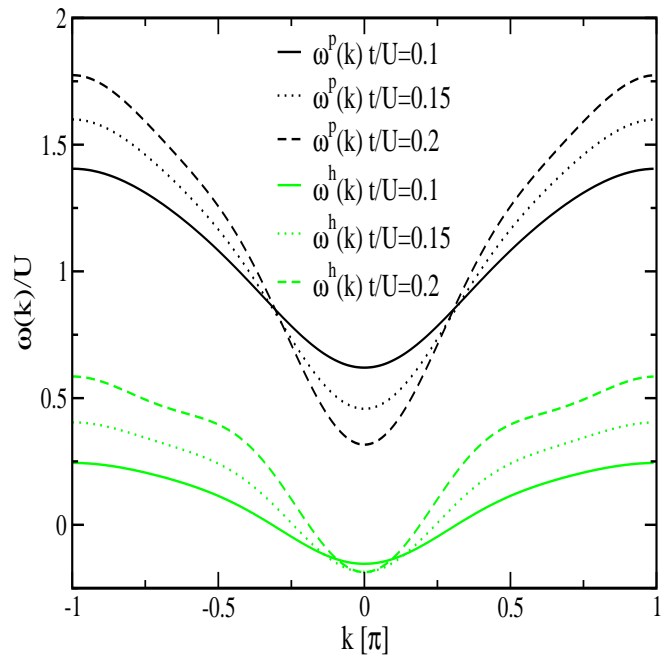}
    \end{center}
    \caption{(color online) The dispersions of a particle $\omega^{\rm p}(k)$ 
      (black) and a hole $\omega^{\rm h}(k)$ (green/grey) for $t=0.1U$ (solid),
      $t=0.15U$ (dotted) and $t=0.2U$ (dashed).}
    \label{fig_disp}
\end{figure}
In Fig.~\ref{fig_disp}, the single-particle dispersion $\omega^{\rm  p}(k)$ 
(one-hole dispersion $\omega^{\rm h}(k)$) is shown as black 
(green or grey, resp.) curves.
Both dispersions increase with $t$; the particle dispersion 
always exceeds the hole dispersion. 
On increasing $t$, the center of the hole dispersion shifts
to higher energies while the center of the particle dispersion 
remains fairly constant in energy.

\section{Observable}
\begin{figure}[t]
    \begin{center}
     \includegraphics[width=\columnwidth]{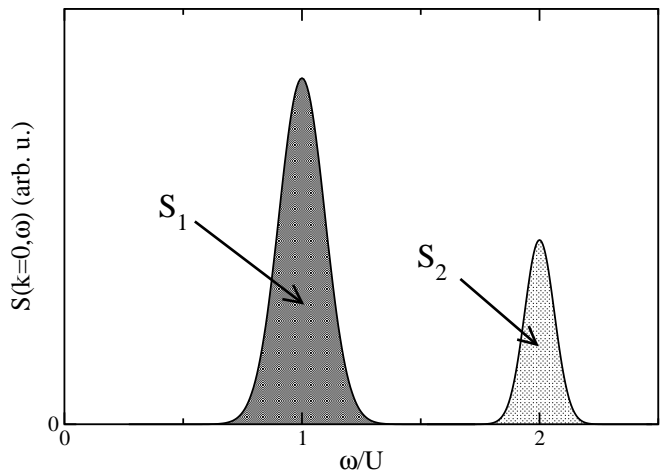}
    \end{center}
    \caption{Sketch of the distribution of spectral weight $S(k=0,\omega)$. 
      The weight centered around $\omega=U$ is given by
      $S_1$, the weight around $\omega=2U$ by $S_2$.}
    \label{fig_sketch}
\end{figure}

At the beginning of the flow ($l=0$), the observable $R$ is proportional to 
$H_t$. The dynamic structure factor $S_R(k,\omega)$ encodes the response of the
system to the application of the observable $R$. The observable $R$ transfers 
no momentum because it is invariant with respect to translations.
Therefore, the Bragg spectroscopy \cite{stofe04,kohl05}  measures the response 
$S_R(k=0,\omega)$  at momentum $k=0$ and  energy $\omega$. 
A sketch of the spectral density $S_R(k=0,\omega)$ for this observable is shown
in Fig.~\ref{fig_sketch}. We assume that the average energy 
is mainly determined by $H_U$ in the MI regime.
Then the first continuum is located at $U$ and a 
second one at $2U$. For small $t/U$ the continua will be well separated. 
The energy-integrated spectral density in the first continuum is the spectral
weight $S_1$. Correspondingly, $S_2$ stands for 
the spectral weight in the second continuum.

To analyze the spectral weights in the effective model obtained by the CUT
the observable $R$ must be transformed as well.
Before the CUT, the observable is $R(l=0)=H_t$. It is a sum of 
local terms 
\begin{equation}
  R(l=0)=H_t= \sum_{{\mathbf r}} 
  b^\dagger_{{\mathbf r}-1/2} b^{\protect\phantom\dagger}_{{\mathbf r}+1/2}
  +b^\dagger_{{\mathbf r}+1/2} b^{\protect\phantom\dagger}_{{\mathbf r}-1/2}\ ,
\end{equation}
where we have rewritten the sum as a sum over the bonds ${\mathbf r}$. The bond
positions are in the centers of two neighboring sites. This notation emphasizes
that the observable acts locally on  bonds. The observable is
transformed by the CUT to
\begin{equation}
\label{eq:Reff}
  {R}^{\rm eff}=R(l=\infty)=\sum_{{\mathbf r}} R(l=\infty,{\mathbf r})\ .
\end{equation}
It is the sum over the transformed local observables $R(l=\infty,{\mathbf r})$ 
which are centered at the bonds ${\mathbf r}$. 

\begin{figure}[h]
    \begin{center}
     \includegraphics[width=\columnwidth]{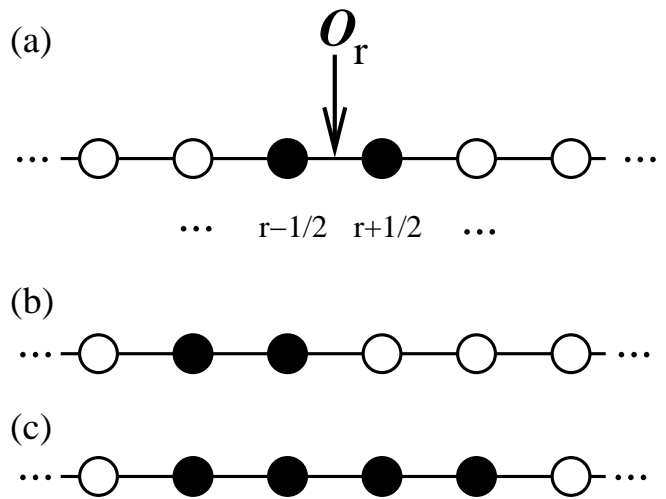}
    \end{center}
    \caption{Terms of the local observable $R(l,{\mathbf r})$. 
      The arrow indicates the bond on
      which a term acts. Filled circles: sites on which a non-trivial
      operator acts; `non-trivial' means that the operator is different from
      the identity. Open circles: sites on which the term under study does not 
      act, i.e., the term acts as identity on these sites.
      (a) At $l=0$ the observable is $H_t$. 
      It is composed of local terms that act only on adjacent sites of the
      lattice. 
      (b) and (c) More complicated terms appear during the flow. The sum
      of the distances of all operators in a term is our measure 
      $r_{\mathcal O}$ for the extension of a term. 
      It is $2$ for (b) and $4$ for (c).  Terms beyond
      a certain extension are omitted.}
    \label{fig_obsaction}
\end{figure}

The local observable is sketched schematically in Fig.~\ref{fig_obsaction}. 
The sites on which the local observable acts are shown as filled circles. 
The state of the sites shown as empty circles is not altered by the observable.
At $l=0$, the observable is $H_t$. It acts only locally on adjacent sites of 
the lattice as shown in Fig.~\ref{fig_obsaction}a. 



We compute the matrix elements of the excitation
operator after the CUT, i.e., of ${R}^{\rm eff}$. This operator
consists of terms of the form 
$c_{(\bar{n};\bar{m})} (h^\dagger)^{n_h}(p^\dagger)^{n_p} (d^\dagger)^{n_d}
h^{m_h}p^{m_p} d^{m_d}$ where we omitted all spatial subscripts
denoting only the powers 
$(\bar{n};\bar{m}) = (n_h\; n_p\; n_d; m_h\; m_p\; m_d)$ of the particular 
operator type. 
For the full expression we refer the reader to Appendix~\ref{sec:app}. 
The coefficient  $c_{(\bar{n};\bar{m})}$ is the corresponding prefactor.
The  spectral weight $I^\text{eff}_{(\bar{n};\bar{m})}$ 
is the integral of $S(k=0,\omega)$ over all frequencies for momentum 
transfer $k=0$. It stands for
 the excitation process starting from the states with $m_d$ double-particles,
$m_p$ particles, and $m_h$ holes and leading to 
the states with $n_d$ double-particles, $n_p$ particles, and $n_h$ holes. 

In the course of the flow, contributions to the observable appear which
do not act on the bond on which the initial observable is centered.
The initial local process spreads out in real space for $l\to \infty$.
Examples are sketched in  Fig.~\ref{fig_obsaction}b-c. 
In order to avoid proliferation, also the terms  in the observable have to be 
truncated. Like for the terms in the Hamiltonian we introduce a measure
for the extension of the terms in the observable. This measure, however,
is slightly different: It is the sum $r_{\mathcal O}$ of the distances 
of all its local creation or annihilation operators  to ${\mathbf r}$.
If the value $r_{\mathcal O}$ of a certain term
exceeds a preset  truncation criterion, this term is neglected.

Figures \ref{fig_obsaction}b-c illustrate the truncation criterion for the
observable. An operator that acts on the sites shown in 
Figure~\ref{fig_obsaction}b meets the truncation criterion for 
$r_{\mathcal O}=3$. It is kept in a $r_{\mathcal O}=3$ calculation. 
An operator acting on the sites in Figure~\ref{fig_obsaction}c 
is kept in the calculation with $r_{\mathcal O}=4$. But it does not meet
the truncation criterion for $r_{\mathcal O}=3$; in a  calculation with 
$r_{\mathcal O}=3$ it is discarded. 

Once the local effective observables $R(l=\infty,{\mathbf r})$ are known,
the total effective expression $R^\text{eff}$ is given by the sum over 
all bonds (\ref{eq:Reff}).

\begin{figure}[thbp]
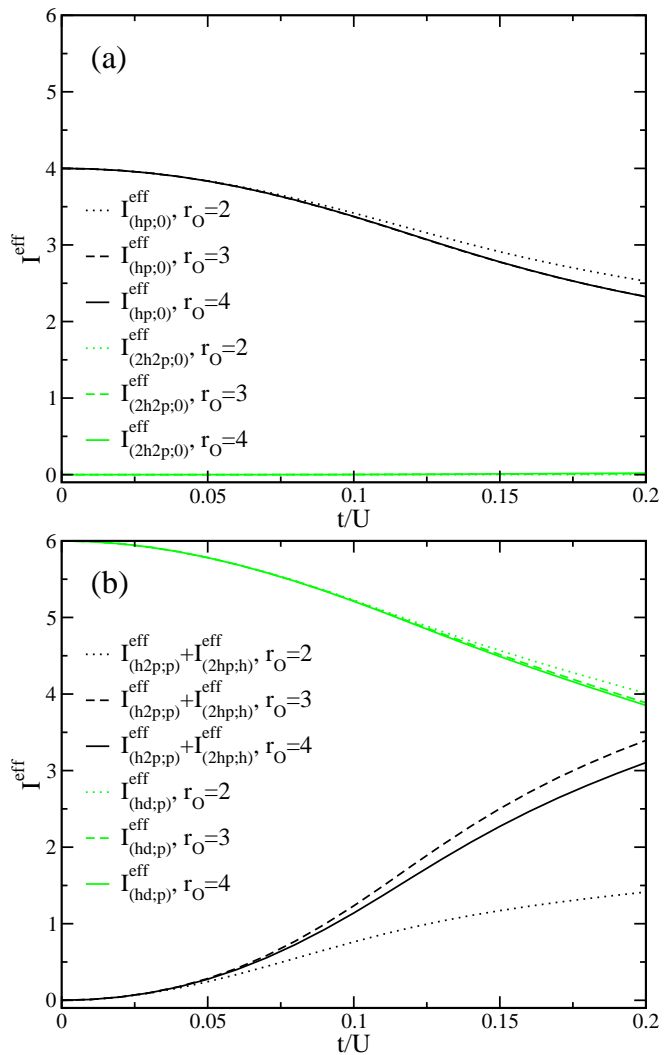

    \begin{center}
      \includegraphics[width=\columnwidth]
      {./fig6a.eps}
      \includegraphics[width=\columnwidth]
      {./fig6b.eps}
    \end{center}
    \caption{(color online) Spectral weights of various processes for 
      truncations $r_{\mathcal O}\in \{2,3,4\}$. 
      (a) 
      $I^\text{eff}_{(\text{hp;0})}$ (black curves) and 
      $I^\text{eff}_{(2\text{h2p};0)}$ (green/grey). 
      (b) 
      $I^\text{eff}_{(\text{h2p;p})}+I^\text{eff}_{(\text{2hp;h})}$ (black)
      and  $I^\text{eff}_{(\text{hd;p})}$ (green/grey)}
    \label{fig_sw}
\end{figure}
The results for the observable truncations
$r_{\mathcal O}\in \{2, 3, 4\}$ are shown in Fig.~\ref{fig_sw}. At zero temperature, only the processes starting from the ground state
are relevant. These processes are the  $(\bar{n};0)$ processes 
which start from the reference state $|\text{ref}\rangle$ because
the CUT is constructed such that
the reference state, which is the vacuum of excitations,
becomes the ground state after the transformation
\cite{knett00a,mielk98,knett03a}.
The  weights of the $(\bar{n};0)$ processes are shown 
in  Fig.~\ref{fig_sw}a. The by far dominant weight is
 $I^\text{eff}_{(\text{hp};0)}$; the process 
$I^\text{eff}_{(\text{2h2p};0)}$ is lower by orders of magnitude.
The agreement between results for various truncations is very good.

The particle-hole pair excited in $I^\text{eff}_{(\text{hp};0)}$ 
has about the energy $U$ for low values of $t/U$, 
while $I^\text{eff}_{(\text{2h2p};0)}$ leads to a response at $2U$.
Hence we find noticeable weight only around $U$ in accordance with
recent quantum Monte Carlo (QMC) data \cite{batro05}.

At finite temperature, excitations  are present before $R^\text{eff}$
is applied. At not too high temperatures, independent particles $p^\dagger$
and holes $h^\dagger$ are the prevailing excitations. Other excitations,
for instance $d^\dagger$ or correlated states $(p^\dagger)^2$
of two $p$-particles, are higher in energy and thus much  less likely.
So the processes starting from a particle or a hole are the important ones
which come into play for $T>0$. Hence we focus on
 $I^\text{eff}_{(\text{h2p;p})}$, $I^\text{eff}_{(\text{2hp;h})}$,
and $I^\text{eff}_{(\text{hd;p})}$. 
These weights are shown in Fig.~\ref{fig_sw}b. The results 
for $I^\text{eff}_{(\text{hd;p})}$ depend only very little on truncation. 
A larger dependence on the truncation is found for
$I^\text{eff}_{(\text{h2p;p})}+I^\text{eff}_{(\text{2hp;h})}$. But the 
agreement is still satisfactory. 

The processes 
$I^\text{eff}_{(\text{h2p;p})}$ and $I^\text{eff}_{(\text{2hp;h})}$
increase the energy by about $U$ because they 
create an additional particle-hole pair. The  
$I^\text{eff}_{(\text{hd;p})}$ process increments the energy
by about $2U$ because a hole and a double-particle is generated.

\section{Approaching Experiment}

Let us address the question what causes the high energy peak in
Refs.~\cite{grein02,stofe04,kohl05}. It was suggested that
certain defects,  namely an adjacent pair of a singly and a doubly
occupied site, are at the origin of the high energy peak.
The weight of such processes for a given particle state is
quantified by $I^\text{eff}_{(\text{hd;p})}$. Its 
relatively high value, see Fig.\ \ref{fig_sw}b, puts the presumption
that such defects cause the peak at $2U$ on a quantitative basis.

\subsection{Zero Temperature}

But what generates such defects? At zero or at very low temperature,
the inhomogeneity of the parabolic trap can imply the existence
of plateaus of various occupations $\langle n\rangle \in \{0,1,2,\ldots\}$
depending on the total filling \cite{batro02,kolla04,batro05}.
In the transition region from one integer value of $\langle n\rangle$
to the next, defects occur which lead to excitations at $2U$.

Yet it is unlikely that this mechanism explains the experimental
finding since the transition region is fairly short at low value of
$T/U$ and $t/U$, i.e., the plateaus prevail. The high energy peak at $2U$, 
however, has less weight by only a factor $2$ to $5$ compared to the weight 
in the low energy peak at $U$ \cite{stofe04}. So we conclude that the 
inhomogeneity of the traps alone  cannot be sufficient to
account for the experimental findings.

\subsection{Finite Temperature}
At higher temperatures, thermal fluctuations are a likely candidate
for the origin of the defects. Hence we estimate their effect in the following
way. Thermally induced triply occupied $d$ states are
neglected because they are very rare due to their high energy of
$\omega^\text{d}(k)-2\mu\approx 2U$ above the vacuum $|\text{ref}\rangle$.
We focus on the particle $p$ and the holes $h$.
The average occupation of these states is estimated by a previously introduced
approximate hardcore boson statistics \cite{troye94}
\begin{subequations}
\label{eq:statistik}
\begin{eqnarray}
 \langle n^{\sigma}_k \rangle &=& 
 \exp\left({-\beta( \omega^{\sigma}(k) -\mu^{\sigma} )}\right)
 \langle n^{\rm vac} \rangle
\\
 \langle n^{\rm vac} \rangle &=& \left(1+z^{\rm p}(\beta) + z^{\rm
 h}(\beta)\right)^{-1}\ ,
\end{eqnarray}
\end{subequations}
where 
$z^\sigma = (2\pi)^{-1}\int_0 ^{2\pi}dk e^{-\beta(\omega^{\sigma}(k)
-\mu^{\sigma})}$, $\mu=\mu^\text{p}=-\mu^\text{h}$. The chemical potential
changes sign for the holes because a hole stands for the absence of
one of the original bosons $b$. Equation (\ref{eq:statistik}) is obtained
from the statistics of bosons without any interaction by correcting globally
for the overcounting of states. The fact that the overcounting is 
remedied on an average level implies that (\ref{eq:statistik}) 
represents only an approximate, classical description \cite{schmi05c}.
The chemical potential $\mu$ is determined self-consistently such that 
as many particles as holes are excited, i.e.,
the average number of bosons $b$ per site remains one. 

\begin{figure}[t]
    \begin{center}
     \includegraphics[width=\columnwidth]{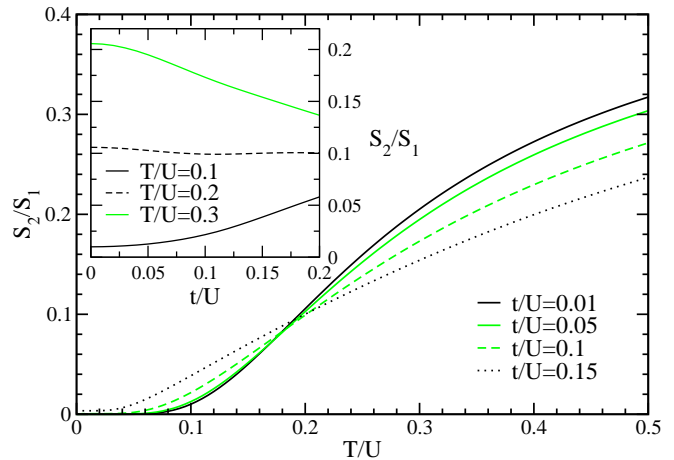}
    \end{center}
    \caption{(color online) Temperature dependence of the relative 
      spectral weight $S_2/S_1$ for various values of $t/U$.
      {\it Inset:} 
      $S_2/S_1$ as function of $t/U$ for various temperatures.}
    \label{fig_sw_T}
\end{figure}
Next, we turn to the computation of spectral weights at finite temperatures.
At not too high temperatures,
the relevant channels are $(\text{hp};0)$, $(\text{h2p;p})$, $(\text{2hp;h})$,
which excite at around $U$ and $(\text{2h2p};0)$, $(\text{hd;p})$,
which excite at around $2U$.
The spectral weights at finite $T$ are calculated by Fermis golden rule
as before. The essential additional ingredient is the probability to
find a particle, or a hole, respectively, to start from and to find
sites which can be excited, i.e., which are not occupied by a particle
or a hole. This leads us to the equations
\begin{subequations}
\begin{eqnarray}
I^{\text{eff},T}_{(\bar{n};0)} 
&=& I^{\rm eff}_{(\bar{n};0)}\langle n^{\rm vac} \rangle^{n_h+n_p+n_d}
\\
 I^{\text{eff},T}_{(\bar{n};\sigma)} &=& 
\frac{\langle n^{\rm vac} \rangle^{n_h+n_p+n_d}}{2\pi}
\int_0^{2\pi} I^{k,{\rm eff}}_{(\bar{n};\sigma)} \langle n^{\sigma}_k\rangle dk
\end{eqnarray}
\end{subequations}
with $\sigma\in\{\text{p,h}\}$. The powers of $\langle n^{\rm vac} \rangle$
account for the probability that the sites to be excited are not blocked
by other excitations; the factor
$\langle n^{\sigma}_k\rangle$ accounts for the probability 
that the excitation necessary for the particular process
is present.
The momentum dependence in $I^{k,{\rm eff}}$ stems from the momentum
dependence of the annihilated particle or hole. It is computed by the
sum of the moduli squared of the 
Fourier transform of the matrix elements $c_{(\bar{n};\bar{m})}$
in the real space coordinate of the annihilated excitation. 

The spectral weight in the high energy peak (at $2U$) relative to the weight in
the low energy peak (at $U$) is given by the ratio $S_2/S_1$ where 
$S_2=I^{\text{eff},T}_{(\text{2h2p};0)}+I^{\text{eff},T}_{(\text{hd;p})}$ and 
$S_1=I^{\text{eff},T}_{(\text{hp};0)}+I^{\text{eff},T}_{(\text{h2p;p})}+
I^{\text{eff},T}_{(\text{2hp;h})}$. Fig.~\ref{fig_sw_T}
displays this key quantity as function of $t/U$ and $T/U$ for 
$r_{\mathcal O}=4$. The difference to the result for $r_{\mathcal O}=3$ is 
less than $0.005$.  

Two regimes can be distinguished. For low values of
$T\lessapprox 0.19U$, the ratio $S_2/S_1$ increases on increasing $t/U$
because the particle gap $\Delta^\text{p}$ decreases so that
$\langle n^\text{p}_{k\approx 0}\rangle$ grows.
For higher values of $T\gtrapprox 0.19U$, the increase of 
$\langle n^\text{p}_{k\approx 0}\rangle$ 
is overcompensated by the decrease of the weights
$I^{\text{eff},T}_{(\text{2h2p};0)}+I^{\text{eff},T}_{(\text{hd;p})}$
and the increase of the weights $I^{\text{eff},T}_{(\text{hp};0)}+
I^{\text{eff},T}_{(\text{h2p;p})}+ I^{\text{eff},T}_{(\text{2hp;h})}$, cf.\
Fig.~\ref{fig_sw}, so that the relative spectral weight
decreases on increasing $t/U$. Around $T=0.19U$, the ratio is fairly
independent of $t/U$.

The experimental value of $S_2/S_1$ \cite{stofe04} is about $0.2-0.5$ 
for small values of $t/U$ ($t\leq 0.03U$).  It increases
on approaching the superfluid phase. Hence, our estimate for
$S_2/S_1$ implies a \emph{significant} temperature $T\approx U/3$
in the MI phase, which was not expected. 
This is the main result of our analysis of the spectral weights.

\section{Discussion}
The analysis of the spectral weights lead us to the conclusion that
the temperature of the Mott-insulating phases must be quite considerable.
At first sight, this comes as a surprise because other experiments
on cold atoms imply very low temperatures, see for instance Ref.\
\onlinecite{pared04}. But the seeming contradiction can be resolved.

One has to consider the entropies involved \cite{blaki04,rey04}.
The entropy per boson $S/N$ in a three-dimensional harmonic trap
is given by $\approx 3.6(1-f_0)$ where $f_0$ is the fraction
of the Bose-Einstein condensate \cite{schmi05c}. 
Adiabatic loading into the optical lattice
keeps $S/N$ constant so that we can estimate the temperature of
the MI phase from the derivative of the free energy \cite{troye94}. 
For $f_0=0.95; 0.9; 0.8$ we obtain  significant temperatures
$T/U=0.12; 0.17; 0.27$ in satisfactory
agreement with the analysis of the spectral weights \cite{schmi05c}.
The analogous estimate in the case of the Tonks-Girardeau limit (large $U$ and
average filling $n$ below unity, see Ref.\ \onlinecite{pared04})
leads to temperatures of the order of the hopping $t$:
$T/t= 0.17; 0.32; 0.61$ (assuming $n=1/2$) \cite{schmi05c}.
These values are again in good agreement with
experiment. 

The physical interpretation is the following.
On approaching the Mott insulator by changing the filling to the
commensurate value $n\to 1$ the temperature has to rise because
the available phase space decreases. For $n<1$ there is phase space 
without occupying any site with two or more bosons $b$ because
one can choose between occupation 0 or 1. But at $n=1$ the state
without any doubly occupied site is unique. Hence no entropy 
can be generated without inducing doubly occupied and empty sites, i.e., 
excitations of $p$- and of $h$-type. This in turn requires that
the temperature is of the order of the gap which agrees well with
our analysis of the spectral weights.

We like to draw the reader's attention to the fact that our result
of a fairly large temperature ($T\approx U/3$) provides also a 
possible explanation why no response at $\approx 2U$  was found
by QMC a low temperatures \cite{batro05}. 
Further investigations will be certainly fruitful, e.g.,
it would be interesting to obtain QMC results for the 
excitation operator $R$ that we used in our present work.
The excitation operator used in Ref.\ \onlinecite{batro05}
vanishes for vanishing momentum so that it is less suited to
describe the experimental Bragg spectroscopy \cite{stofe04}.

In the attempt to provide quantitive numbers for the temperature
in the MI bose systems one must be cautious because the bosonic systems
are shaken fairly strongly in experiment \cite{stofe04,kohl05}. 
Though it was ascertained that the experiment was conducted
in the linear regime it might be that the systems are heated by
the probe procedure so that the temperatures seen in the 
spectroscopic investigations are higher than those seen by other
experimental investigations.

It would be rewarding to clarify the precision of the
 spectroscopic investigations. Then the pronounced $T$ dependence of the
relative spectral weight in Fig.\ \ref{fig_sw_T} could be used as
a thermometer for Mott-insulating bosons in optical lattices.

In summary, we have studied spectral properties of bosons in one-dimensional
optical lattices using particle-conserving continuous
unitary transformations (CUTs). At $T=0$ and for small $t/U$ 
spectral weight is only present at energies $\approx U$. 
Recent experimental peaks at $\approx 2U$ \cite{stofe04} can
be explained assuming $T\approx U/3$. Our results suggest to investigate the
effects of finite $T$ on bosons in optical lattices much more thoroughly.

\begin{acknowledgments}
We thank T. St\"oferle for providing the experimental data.
Fruitful discussions are acknowledged with A. L\"auchli, 
T. St\"oferle, I. Bloch,
S. Dusuel, D. Khomskii, and E. M\"uller-Hartmann. 
This work was supported by the DFG via SFB 608 and  SP 1073.
\end{acknowledgments}

\bigskip

\begin{appendix}
\section{Explicit expression for the observable}\label{sec:app}
The local observable depending on the flow parameter $l$ reads  
\begin{eqnarray}
  \label{eq:bose:obsformel2}
  R(l,\mathbf r)&=&\sum_{\{i, i^\prime, j, j^\prime, k,k^\prime\}}
c(l,\{i, i^\prime, j, j^\prime, k,k^\prime\})\times\nonumber\\
&&\quad h_{i_1+r}^{\dagger}\cdot ... \cdot h_{i_{n_h}+r}^{\dagger}  h_{i^\prime_1+r}^{\phantom{\dagger}} \cdot ... \cdot h_{i^\prime_{m_h}+r}^{\phantom{\dagger}}\times\nonumber\\
&&\quad p_{j_1+r}^{\dagger}\cdot ... \cdot p_{j_{n_p}+r}^{\dagger}  p_{j^\prime_1+r}^{\phantom{\dagger}} \cdot ... \cdot p_{j^\prime_{m_p}+r}^{\phantom{\dagger}}\times\nonumber\\
&&\quad d_{k_1+r}^{\dagger}\cdot ... \cdot d_{k_{n_d}+r}^{\dagger}
d_{k^\prime_1+r}^{\phantom{\dagger}} \cdot ... \cdot
d_{k^\prime_{m_d}+r}^{\phantom{\dagger}}\nonumber. \\
\end{eqnarray}
The numbers $(n_h\; n_p\; n_d; m_h\; m_p\; m_d)$ are defined as the number of 
operators involved in a term in Eq.~\ref{eq:bose:obsformel2}. The number of
creation operators $h^\dagger$, $p^\dagger$, and $d^\dagger$ is 
given by  $n_h$, $n_p$, and $n_d$, respectively. 
The number of annihilation operators $h$, $p$, and $d$ 
is given by $m_h$, $m_p$, and $m_d$, respectively. 
A set of these six numbers defines the type
$(\bar{n};\bar{m})$ of a process. 
The variables $\{i, i^\prime, j, j^\prime, k,k^\prime\}$
are multi-indices, e.\ g.\ $i=\{i_1,...,i_{n_h}\}$ which give the position of
the operator. The coefficients $c(l,\{i, i^\prime, j, j^\prime, k,k^\prime\})$
keep track of the amplitudes of these processes during the flow. Their value
at $l=\infty$ defines the effective observable $R^\text{eff}$. 

The total observable is the sum 
\begin{equation}
  R(l)= \sum_{{\mathbf r}} R(l,{\mathbf r}).
\end{equation}
No phase factors occur so that the observable is invariant with respect
to translations. Hence no  momentum transfer takes place and the
response at $k=0$ is the relevant one. 
\end{appendix}


\end{document}